\documentstyle[epsf,floats,aps,prl]{revtex}
\begin{document}
\draft
\twocolumn[\hsize\textwidth\columnwidth\hsize\csname @twocolumnfalse\endcsname
\title{Hall Resistivity and Dephasing in the Quantum Hall Insulator}
\author{Leonid P.~Pryadko$^1$ and Assa Auerbach$^{2}$}
\address{
$^1$Department of Physics, University of California, Los Angeles, CA\\
$^{2}$Department of Physics, Stanford University, CA 94305,  \\
and  Department of Physics, Technion, Haifa 32000, Israel.} 
\date{\today }
\maketitle
\begin{abstract}
  The longstanding problem of the Hall resistivity $\rho_{xy}$ in the
  Hall insulator phase is addressed using four-lead Chalker-Coddington
  networks.  Electron interaction effects are introduced via a finite
  dephasing length.  In the quantum coherent regime, we find that
  $\rho_{xy}$ scales with the longitudinal resistivity $\rho_{xx}$,
  and they both {\em diverge\/} exponentially with dephasing length.
  In the Ohmic limit, (dephasing length shorter than Hall puddles'
  size), $\rho_{xy}$ remains quantized and independent of $\rho_{xx}$.
  This suggests a new experimental probe for dephasing processes.
\end{abstract}
\vskip1pc]
\narrowtext

The ideal Quantum Hall (QH) effect can be defined as the {\em
  simultaneous\/} quantization of Hall conductance and vanishing
longitudinal conductance:
\begin{equation}
  \label{strong-QHE}
  G_{xy}=\nu\,  e^2/h,\quad
  G_{xx} =0, 
\end{equation}
where $\nu$, the filling factor, is some integer or odd denominator
fraction.  Consider a perfect QH sample with four Ohmic leads.  By
Kirchoff's laws, it is easy to see that while the {\em measured\/}
$G_{xy}, G_{xx}$ and $R_{xx}$ depend on the lead resistances, the Hall
resistance is independent of them, and is precisely quantized at
\begin{equation}
  R_{xy}=  h/\nu e^2 . 
  \label{weak-QHE}
\end{equation}
This robustness of $R_{xy}$ generalizes to larger circuits, as proven
by Shimshoni and Auerbach\cite{Shimshoni-97} for the two dimensional
Ohmic puddle network model.  However, as we shall see below, the
quantum interference between different tunnel junctions (absent
for the Ohmic transport) spoils the quantization of
$R_{xy}$\cite{Pryadko-98}.

Quantum interference also drives localization and related $T=0$
QH to insulator
transition\cite{Pruisken-88,Pruisken-localization}.  This has been
confirmed by explicit calculations of the localization length
exponent, in quasiclassical approximation\cite{Milnikov-88} and
numerically\cite{Chalker-88,Lee-93} on Chalker-Coddington (CC)
networks.

There is no consensus however, about interference effects on the Hall
resistivity; its value in the insulating phase is still controversial.
Several theories expect it to be finite\cite{Zhang-92,Kivelson-92}
$\rho_{xy}^{T=0} < \infty$, or even quantized\cite{Dykhne-94,Ruzin-95}
$\rho_{xy}^{T=0}=h/e^2$ (``semicircle law'').  Experimental
observations also vary: some data can be fit by the Drude
form\cite{Goldman-88,Jiang-93} $\rho_{xy}\propto B$, while others see
a weaker magnetic field ($B$) dependence, with $\rho_{xy}$ nearly
quantized\cite{Shahar-96}.

In contrast, Entin-Wohlman~{\em et.\ al.\/}\cite{Entin-Wohlman-95}
pointed out that $\rho_{xy}^{T=0}$ is not a self-averaging quantity in
the insulator.  Using a model of local hopping in external magnetic
field, they concluded that %$\lim_{T\to 0} \rho_{xy}=\infty$.
$\rho_{xy}$ diverges in the limit $T\to 0$.  Unfortunately, the
applicability of this model to QH systems is somewhat questionable.
\begin{figure}[htb]
  \epsfxsize=0.55\columnwidth\centerline{\epsfbox{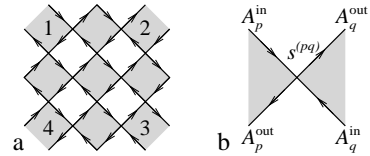}}
  \vspace{0.02in}
  \caption{
    (a) Four-terminal $L=6$ CC network.
    %% which defines the transmission matrix $T$ of
    %% Eq.~(\protect\ref{Rab}).
    QH puddles are shaded.  Edge currents with amplitudes $A^\alpha_p$
    propagate along the arrows.  (b)~The incoming and outgoing
    amplitudes at the tunnel junction between puddles $p$ and $q$ are
    related by the scattering matrix~(\protect\ref{Spq}).}
  \label{fig:Icr}
\end{figure}

The purpose of this paper is to determine the low temperature Hall
resistivity in the magnetic field driven insulating phase.  Differing
from a previous calculation\cite{Ruzin-95}, we directly compute the
four-terminal resistances $R_{\alpha\beta}$, $\alpha,\beta\!=\!x,y$
for random CC networks using the Landauer-B\"uttiker quantum transport
theory.  Distributions of $R_{\alpha\beta}$ are obtained for different
linear system sizes $L$ and magnetic field parameters $\theta$,
defined below.

Our key result is that at large $L$, the Hall resistance
$R_{xy}(\theta,L)$ of the CC network changes from the
asymptotically approaching quantized value $h/\nu e^2$ in the QH phase
to an {\em exponential divergence\/} in the insulator\cite{ShengWeng}.
Second, effects of inelastic scattering due to electron-electron and
electron-phonon interactions are introduced via a phenomenological
{\em dephasing length\/} $l_\varphi$\cite{Pruisken-88,Sondhi-97} (for
convenience, we measure all physical length in the units of typical
puddle size $l_V$).  For $l_\varphi$ shorter then the puddles' size
(phase-incoherent limit) we map the chiral network to the Ohmic puddle
network model\cite{Shimshoni-97}, which has a quantized Hall
resistance.  In general, the macroscopic resistivity is approximately
given by an Ohmic resistor network formed by elements of size
$L=l_\varphi$ with the computed distributions of $R_{\alpha\beta}$.
The Ohmic network problem is solved using percolation theory following
Ambegaokar, Halperin and Langer\cite{Ambegaokar-71} (AHL).  This
approach, justified by the wide distribution of $R_{xx}$, yields the
ensemble of elements which determine the Hall voltage.  The
macroscopic Hall resistivity $\rho_{xy}$ is obtained as an average
over this ensemble.  We show that the divergence of both Hall and
longitudinal macroscopic resistivities in the insulator is consistent
with one length scale $\xi(\theta)$.
We conclude with the suggestion that $\rho_{xy}(T)$ could serve as an
experimental probe of the dephasing length.
 
{\em Four-terminal CC network.\/} Consider a square CC network in
Fig.~\ref{fig:Icr}, with corner puddles serving as external leads.
Carriers, described at a given energy by complex amplitudes
$\{A_{p}^\alpha\}$, where $\alpha,p$ are the edge and puddle labels
respectively, can propagate only in one direction around the puddles,
as indicated by arrows.  The corresponding currents are just
$I_{p}^\alpha= |A_{p}^\alpha|^2$.  Edge amplitudes across each tunnel
junction $pq$ are related by a unitary scattering matrix $s^{(pq)}$,
\begin{equation}
  A^{\rm out}_r\!\!=\!\!\!\! \sum_{r'=p,q}\!\!\!
  s^{(pq)}_{rr'}\,e^{i\phi_{r'}^{\rm in}} A^{\rm in}_{r'}, \;\,
  s^{(pq)}=\! 
    { \;\;\cos\theta_{pq}\;\; \sin\theta_{pq}
      \choose -\sin\theta_{pq} \;\cos\theta_{pq}},
  \label{Spq}
\end{equation}
where $\theta_{pq} \in [\theta+\delta,\theta-\delta]$, and edge phases
$\phi_{r}^{\alpha}\in [0,2\pi)$ are independent random variables; we
have chosen\cite{Lee-93} $\delta=0$ to reduce the statistical errors.
The magnetic field parameter $\theta$ can be tuned across the
QH--insulator transition at $\theta_c=\pi/4$.  In the QH regime
$\theta < \theta_c$ the neighboring leads are connected by
highly-transmitting edge states, while the insulating phase
$\theta>\theta_c$ consists of weakly-connected puddles without global
edge states.
%% AA
Near the transition, $\theta-\theta_c \propto B-B_c$.  Away from the
transition, tunneling across saddle points yields asymptotically
$\tan(\theta) \propto e^{-A(B-B_c)}$\cite{Shimshoni-98}.
%% END

The scattering relationships provide a set of linear equations for all
$A^\alpha_p$ which depends on the incoming amplitudes $A^{\rm in}_i$
at the external leads $i=1$--$4$.  The global scattering matrix
$S_{ij}[\theta_r,\phi_r^\alpha]$ relates the 
amplitudes at the external leads, $A^{\rm out}_i =\sum_{j=1}^4
S_{ij}\, A^{\rm in}_j$, so that the current transmission matrix
$T_{ij}=|S_{ij}|^2$ defines the probability of scattering from
incoming channel $i$ to outgoing channel $j$.  We assume that the
external leads, also formed by an incompressible QH liquid, are in
equilibrium, which implies that the chemical potentials of their edges
are\cite{Beenakker}
\begin{equation}
  \mu_i^\alpha = \left({h/\nu e}\right) I_i^\alpha.
  \label{edge-states}
\end{equation}
The resistance tensor is now given by a compactified version of the
B\"uttiker-Landauer formula\cite{Buttiker-86},
\begin{eqnarray}
  \label{Rab}
  {\tilde R}_{\alpha\beta}&=& \left({h/ \nu e^2}\right)\,
  \Lambda^t_{\alpha}
  (1+T)\,P \left( P\,(1- T)\,P\right)^{-1}\, P\Lambda_{\beta}
  ,\\
  &\quad&\Lambda^t_{x}=(1,0,-1,0),\quad\Lambda^t_{y}= (0,1,0,-1),
  \nonumber
\end{eqnarray}
where the vectors $\Lambda_\alpha^t$ are transposed $\Lambda_\alpha$,
and the operator $P$ projects out the zero eigenvector $(1,1,1,1)$ of
the matrix $1-T$ (matrix $S$ is unitary).  It is easy to show that the
field-antisymmetrized Hall resistance is $R_{xy}=({\tilde
  R}_{xy}-{\tilde R}_{yx})/2$.

%% LP
Numerically, we calculated the $L\times L$ transfer matrix relating
the amplitudes on $L$ open wires at the left and right sides of the
sample (see Fig.~\ref{fig:Icr}).  We used the transfer-matrix
technique with intermediate orthogonalizations\cite{Wang-98} to reduce
numerical errors for large systems.  Then the closed edges were
``linked'' by eliminating the corresponding pairs of incoming and
outgoing amplitudes.
%% END
This resulted in a reduced $4\times4$ transfer matrix, with which the
full scattering matrix $S_{ij}$ was computed.  The numerical accuracy
was controlled by checking the unitarity of $S_{ij}$, and by test runs
at quadruple accuracy.

To obtain the distributions of $R_{xx}$ and $R_{xy}$ for different
values of $\theta$ and the system size $L$, we repeated each
calculation up to $10^6$ times, taking $({\tilde R}_{xx}\,,{\tilde
  R_{yy}})$ as two different members which have the same $R_{xy}$.  In
the insulating phase $\theta>\theta_c$, $R_{xx}$ has a very wide,
resembling log-normal, distribution over several decades, as generally
expected in a localized insulator phase\cite{Entin-Wohlman-95}.  The
Hall resistance $R_{xy}$ is seen to have a much narrower distribution,
but with mean and variance also increasing exponentially with the size
of the system in the insulating phase.

{\em Phase-incoherent network model}.  In the opposite limit we
consider a network where inelastic mechanisms completely destroy
quantum interference between different tunnel junctions.  In this case
different tunneling events happen independently,
%%Experimentally, measurements are performed at a finite temperature,
%%usually in macroscopic samples, larger then the phase breaking length.
%%Carriers diffuse through the sample coherently until the occurrence of
%%a phase-breaking event.  If the rate of these events is high enough,
%%the interference is altogether suppressed,
and the incoming and outgoing currents are related at each junction,
\begin{equation}
  I^{\rm out}_r=\sum\nolimits_{r'=p,q}\,|s_{rr'}^{pq}|^2\, I^{\rm
    in}_{r'}, 
  \label{Scl}
\end{equation}
where the tunneling matrix elements $s^{(pq)}$ are given by
Eq.~(\ref{Spq}).  According to Eq.~(\ref{edge-states}), local chemical
potentials at each edge are proportional to the corresponding
currents.  Thus, the dissipative and Hall voltages at each junction
are related to the tunneling current $I_{pq}$ as
\begin{equation}
  V^{\rm dis}_{pq}=R_{pq}\, I_{pq},\quad %
  V^H_{pq}= {\rm sign}(B)\,\left({h/ \nu e^2}\right)\, I_{pq},
  \label{voltages}
\end{equation}
where $R_{pq} = (h/\nu\,e^2) \cot^2\theta_{pq}$ is symmetric under
reversal of magnetic field.  These local relations reduce the network
model to the Ohmic puddle network model of Ref.\CITE{Shimshoni-97},
for which $R_{xy}(l_\varphi\!\leq\!l_V)= (h/\nu e^2)$ for any
realization of $\{\theta,\phi\}$.  This model equally applies to
integer and fractional QH regimes.  Eqs.~(\ref{voltages}) also imply
that the phase-incoherent network model has an exact current--voltage
duality\cite{Dykhne-94,Ruzin-95,Pryadko-98}, which interchanges the QH
and insulating regions and simultaneously inverts the resistance
$R_{pq}$ associated with each junction.  This duality also inverts the
macroscopic dissipative resistance of the system, determined,
according to the AHL percolation argument\cite{Ambegaokar-71}, by the
median of the distribution of $R_{pq}$.
%%It is also easy to see that this
%%model can be made to be self dual
%%\cite{Dykhne-94,Ruzin-95,Pryadko-98}, as follows.  By using wide leads
%%(of half the system width), an interchange of QH and insulating
%%regions with an inversion of each resistance $R_{pq}$ is an identity
%%transformation. Note that

To describe the regime of intermediate dephasing length,
$l_V<l_\varphi<\infty$, where $l_V$ is a typical puddle size, we use
finite-size four-lead CC networks. In the simulation, the phase
breaking occurs only at the leads, and we {\em define} $L=l_\varphi$.
Square blocks of linear sizes $l_\varphi$ or larger connect as Ohmic
resistors, with the resistances $R_{\alpha\beta}(L=l_\varphi)$ chosen
from the numerically determined distribution.  Ignoring relatively
small Hall voltages at this stage, we notice that $R_{xx}$ is
exponentially widely distributed.  So, we can again follow AHL and
insert all resistors in increasing order of $R_{xx}$ until the
percolation threshold is reached.  Here we assume a percolation
threshold of 50\%, applicable for self-dual lattices.  Thus the last
resistors to connect the percolating cluster (PC) have the median value
$R^{\rm med}_{xx}$.  Since all higher resistors $R_{xx}>R_{xx}^{\rm
  med}$ transport negligibly little current, they can be discarded.
By AHL, and subsequent numerical confirmations\cite{Ambegaokar-71},
the macroscopic resistivity of the network is simply $\rho_{xx}=R^{\rm
  med}_{xx}(l_\varphi)$.

The Hall resistivity $\rho_{xy}$ can now be determined as the weighted
average over the PC.  Since numerically, $|R_{xy}|\ll R_{xx}$, and we
draw no current from the Hall leads, it is safe to assume that the
Hall voltage fluctuations across the PC are averaged out by secondary
local currents, and the macroscopic Hall resistivity can be estimated
by the average $\rho^{< med}_{xy}= \langle R_{xy}\rangle_{R_{xx} \le
  R^{\rm med}}$.  As an upper bound, we also calculate the average
Hall resistance of the necks of the PC given by $\rho^{\rm med}_{xy}=
\langle R_{xy}\rangle_{R_{xx} \approx R^{\rm med}}$, which gives
similar results as shown below.
%%% The Hall resistivity $\rho_{xy}$ can now be determined as the
%%% weighted average over the PC.  Although we have not established the
%%% appropriate distribution function on the PC, we tried two limiting
%%% cases: (i) averaging over {\em all} resistors in the PC $\rho^{<
%%%   med}_{xy}= \langle R_{xy}\rangle_{R_{xx} \le R^{\rm med}}$, and
%%% (ii) averaging only over ``hot spot'' resistors in the PC
%%% $\rho^{\rm med}_{xy}= \langle R_{xy}\rangle_{R_{xx} \approx R^{\rm
%%%     med}}$. We find similar behavior for the two cases (see Fig.
%%% \ref{fig:lscale}).

\begin{figure}[htb]
  \epsfxsize=\columnwidth%
  \centerline{\epsfbox{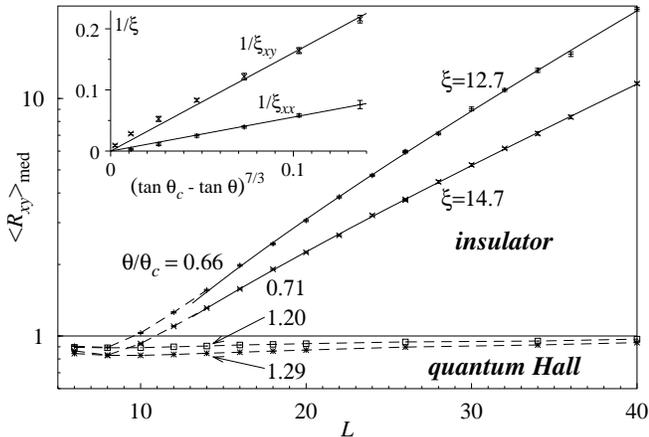}}%
%  \vspace{0.05in}
  \caption{QH to insulator transition
    in finite size scaling.  Solid lines are the best fits of
    Eq.~(\protect\ref{eq:corr-length}), and dashed lines are guide to
    the eye.  Inset: The resulting inverse correlation lengths are
    consistent with the critical exponent $\nu=7/3$}
  \label{fig:lscale}
\end{figure}
\begin{figure}[htb]
  %%\centerline{\psfig{figure=fig-rxxrxy.eps,width=3in}}
  \epsfxsize=0.80\columnwidth\centerline{\epsfbox{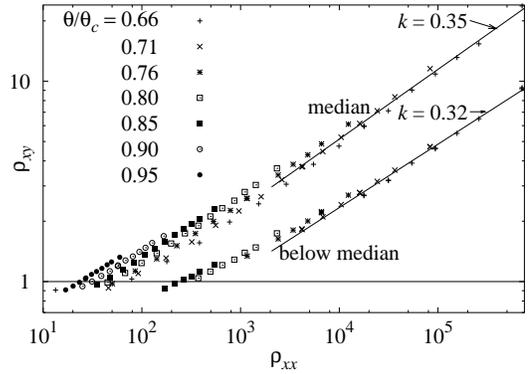}}
  \vspace{0.05in}
  \caption{Asymptotic correlation between $\rho_{xx}$ and $\rho_{xy}$
    determined by two averaging procedures (see discussion in text).
    $k$ is the slope on the log-log plot.  }
  \label{fig:rxxvrxy}
\end{figure}

{\em Results.\/} In the localized regime, $\theta> \theta_c$, both
$\rho_{xx}$\cite{Lee-93} and $\rho_{xy}$ diverge exponentially at
large $l_\varphi$.  Both resistivities fit very well [see
  Fig.~\ref{fig:lscale} for the fit of $\rho_{xy}(L)$] to an
exponential scaling function with finite-size correction
\begin{equation}
  \label{eq:corr-length}
  \rho(L) \sim A_\theta L^\gamma \exp\left[ L/\xi(\theta)\right],
\end{equation}
where $\gamma$ is a $\theta$-independent exponent determined by the
geometry of the system, and $\xi(\theta)$ is the localization length
(both quantities are defined separately for $\rho_{xx}$ and
$\rho_{xy}$).  As illustrated in the inset in Fig.~\ref{fig:lscale},
the divergence of both correlation lengths as $\theta\to \theta_c$ is
consistent with $\xi(\theta)\sim (\theta_c - \theta)^{-7/3}$.  This
form for $\xi_{xx}$ is in agreement with previous
studies\cite{Chalker-88,Lee-93}.  The fact that the correlation length
for $\rho_{xy}$ also diverges with the same exponent implies that the
transition is actually governed by a single length scale
$\xi(\theta)$, and asymptotically $\rho_{xy}\sim
\left(\rho_{xx}\right)^k\to\infty$ with $k\approx 0.32-0.35$
(Fig.~\ref{fig:rxxvrxy}).  This divergence, one of the primary results
of this paper, contradicts the
expectations\cite{Zhang-92,Kivelson-92,Dykhne-94,Ruzin-95} of finite
Hall resistivity in the insulator at $l_\varphi\to \infty$.  Our
results agree with the conclusion of Ref.~\CITE{Entin-Wohlman-95},
although here we have used a different model, applicable for transport
in quantizing magnetic fields.

Fig.~\ref{fig:lscale} also shows $\rho_{xy}$ in the metallic QH phase
for $\theta > \theta_c$. In this phase, as expected, $\rho_{xy}(L)$
approaches the quantized value at large system sizes.  We did not
attempt to determine the corresponding correlation length since our
geometry has narrow leads (see Fig.~\ref{fig:Icr}).  We did, however,
check that our results are not limited to systems with narrow leads by
making a limited set of runs for a special self-dual network geometry,
which remains identical under the interchange of QH and insulating
regions and the replacement $\theta\to2\theta_c-\theta$.

We also attempted to suppress the quantum interference by calculating
the ensemble averaged $\langle T_{ij}\rangle$, which is formally
equivalent to considering the same non-interacting system at very high
temperatures.  This averaging resulted in both Hall and longitudinal
resistance of insulating phase much smaller then the average quantum
values.  However, the specific values of these resistances differed
significantly from the values obtained numerically for the
phase-incoherent networks of identical geometry; particularly, the
Hall resistance was not quantized, with the offset increasing into the
insulating phase.  This demonstrates that at $T>0$, without inelastic
scattering quantum interference cannot be completely suppressed.

In the fractional Hall effect regime, electron-electron interaction 
effects are primarily threefold:
(i)~Stabilization of fractional $\nu < 1$ QH phases in the puddles,
(ii)~Renormalization of inter-puddle electron tunneling rates, and 
(iii)~Dephasing of charge carriers by inelastic scattering.  At low
temperatures the second effect is expected to be finite, since
infrared divergence of the tunneling amplitudes is cut off by the finite
size of the puddles.  In this regime, the effective edge state
transport theory for the fractional and integer cases is
identical\cite{Pryadko-98}, up to factors of $\nu$ in the chemical
potential relations~(\ref{edge-states}).  Thus it is legitimate
to use the CC model as a model of coherent quantum transport
on edges of fractional QH puddles.  However 
the 
dephasing mechanism and determination of 
$l_\varphi$ remains an interesting open problem, 
for both integer and fractional
cases.

{\em Quantum critical point and dephasing}.  
%% LP
In a vicinity of the true
quantum critical point, the effective dephasing length\cite{Sondhi-97}
should diverge at small temperatures as
\begin{equation}
  l_\varphi(T)\sim T^{-1/z}. 
  \label{lphi-p}
\end{equation}
Experimentally, 
%% END
resistivity and non linear resistivity data at transitions between
plateaux have been collapsed onto universal curves using $z\approx
1.0$ (and independently determined $\nu_\xi\approx
2.4$)\cite{Wei-88,Jiang-93,Shahar-97b}.  Based on our results, at 
a quantum critical point, and in the absence of additional
phase-breaking mechanisms, one would also expect the Hall resistivity
to diverge as
\begin{equation}
  \rho_{xy} \sim (h/ \nu e^2)\, %
  \exp\left[ k T^{- 1/z} (B-B_c)^{7/3}\right].
\end{equation}
Experiments, however, have reported a constant or weakly $B$-dependent
Hall resistivity on the insulating side of the transition.  For
similar samples, resistivity saturation at low temperatures has been
reported\cite{Shahar-97c}.  Evidently, {\em both}
effects are inconsistent with a true zero-temperature quantum Hall to
insulator transition,  characterized
by a diverging dephasing length~(\ref{lphi-p}).

We conclude that a nearly quantized Hall resistivity indicates a
strongly dephased regime where, {\em i.e.},\ $l_\varphi \le l_V$.
%, and
%$l_V$ is the inter-junction distance
In contrast, the longitudinal resistivity in this regime is expected
to diverge at large fields as\cite{Shimshoni-98},
\begin{equation}
  \rho_{xx}(B) \sim \left({h/ e^2}\right) %
  \exp\left[ \nu\,\left(l_V/ l\right)^2 \left({B/B_c-1}\right)\right],
  \label{Rxx-pnm}
\end{equation}
where $l$ is the Landau length.  This expression allows an independent
estimation of $l_V$, and an upper bound on the dephasing length in the
limit of small temperatures.  Experimentally, the interplay between
$l_V$, the scale of the long-range potential fluctuations, and the
phase breaking length $l_\varphi$ was noticed in Ref.\CITE{Wei-92}.
 
It is not yet clear what mechanism can explain zero temperature finite
resistivity in disordered QH systems, or why dephasing seems to be
more pronounced in some particular experiments.  One possible source
might be a coupling of the edge excitations to nearby domains of
compressible $\nu=1/2$ phase.

{\em Acknowledgments\/}: Useful discussions with S.~A.~Kivelson,
E.~Shimshoni, M.~Hilke, D.~Shahar, S-C.~Zhang and C.~M.~Marcus are
gratefully acknowledged.  AA acknowledges the hospitality of the
Physics Department, Stanford University, and a grant from the Israel
Science Foundation.

\end{document}